\documentclass[aps,prl,reprint,superscriptaddress]{revtex4-2}
\usepackage{amsmath,amssymb,amsfonts}
\usepackage{graphicx}
\usepackage{natbib}
\usepackage{color}
\usepackage{cancel}
\usepackage{ragged2e}
\usepackage[T1]{fontenc} 
\usepackage{tangocolors}
\usepackage[dvipsnames]{xcolor}
\usepackage{amsmath}
\usepackage{tikz}
\usetikzlibrary{arrows, decorations,backgrounds, patterns}
\usetikzlibrary{decorations.pathreplacing ,decorations.markings}
\usepackage{amssymb}\usepackage{amsfonts}
\usetikzlibrary{decorations.pathmorphing,backgrounds,shapes,arrows,shadows}
\usetikzlibrary{patterns.meta}
\usetikzlibrary{patterns,decorations.pathmorphing}
\tikzset{
    snake it/.style={decorate, decoration=snake}
}
\usepackage{pgfplots}
\pgfplotsset{compat=1.11}
\usepgfplotslibrary{fillbetween}
\usetikzlibrary{intersections}
\pgfdeclarelayer{bg}
\pgfsetlayers{bg,main}
\tikzset{zigzag/.style={decorate,decoration=zigzag}}
\tikzset{snake it/.style={decorate, decoration=snake}}
\makeatletter
\def\@hex@@Hex#1%
 {\if a#1A\else \if b#1B\else \if c#1C\else \if d#1D\else
  \if e#1E\else \if f#1F\else #1\fi\fi\fi\fi\fi\fi \@hex@Hex}
\makeatother

\usetikzlibrary{arrows.meta}
\tikzset{ mid arrow/.style={ postaction={decorate}, decoration={ markings, mark=at position 0.6 with {\arrow{#1}} } } }
\tikzset{>={Latex[width=2mm,length=2mm]}}

\usepackage[all]{xy}
\DeclareFontEncoding{LS2}{}{\noaccents@}
\DeclareFontSubstitution{LS2}{stix}{m}{n}

\DeclareSymbolFont{integrals}{LS2}{stixcal}{m}{n}

\DeclareMathSymbol{\ointupbig}{\mathop}{integrals}{"E8}
\DeclareMathSymbol{\ointupsmall}{\mathop}{integrals}{"B2}

\usepackage[percent]{overpic}
\usepackage{slashed}
\usepackage{wrapfig}
\usepackage{tabu}
\usepackage{diagbox}
\usepackage{mathrsfs,amsmath,amssymb,amsthm,amsfonts,tikz,graphicx,accents,hyperref, color}
\usepackage{dsfont,epiolmec, latexsym, stmaryrd, comment}
\usepackage{slashed,ccaption}
\usepackage{mathrsfs, calligra}
\usepackage{leftidx}
\usepackage{import}
\usepackage{multirow}
\usepackage{amsfonts}
\usepackage{pifont}
\usepackage{tabularx}
\usepackage{cancel}
\usepackage[utf8]{inputenc}
\usetikzlibrary{intersections,calc}
\usepackage{ifthen}
\usepackage{amsmath}
\usepackage{cancel}
\usepackage{caption} 
\usepackage{subcaption}

\usetikzlibrary{patterns.meta}
\usepackage{array}
\hypersetup{ linktoc=all,
    colorlinks, linkcolor={palatinateblue},
    citecolor={brightpink}, urlcolor={amaranth}
}

\graphicspath{{Images/}}

\definecolor{rosy}{RGB}{230,235,252}
\definecolor{myframetitle}{RGB}{90,89,170}
\definecolor{myblocktitle}{RGB}{140,185,249}
\definecolor{mytitle}{RGB}{10,80,26}

\definecolor{darkgreen}{RGB}{27,130,45}
\definecolor{darkblue}{rgb}{0,0,0.3}
\definecolor{darkred}{rgb}{0.7,0,0}

\definecolor{light gray}{RGB}{220,220,220}
\definecolor{dark purple}{RGB}{108,0,217}
\definecolor{pink}{RGB}{190,20,100}
\definecolor{orang}{RGB}{193,63,0}
\definecolor{green}{RGB}{11,98,17}
\definecolor{darkpink}{RGB}{153,0,76}
\definecolor{bluegreen}{RGB}{0,102,102}
\definecolor{greenlagan}{RGB}{0,102,0}
\definecolor{redgreen}{RGB}{102,102,0}
\definecolor{Redgreen}{RGB}{153,76,0}
\definecolor{vividviolet}{rgb}{0.62, 0.0, 1.0}
\definecolor{amaranth}{rgb}{0.9, 0.17, 0.31}
\definecolor{palatinateblue}{rgb}{0.15, 0.23, 0.89}
\definecolor{brightpink}{rgb}{1.0, 0.0, 0.5}
\definecolor{cornflowerblue}{rgb}{0.39, 0.58, 0.93}
\definecolor{deepcarminepink}{rgb}{0.94, 0.19, 0.22}
\definecolor{radicalred}{rgb}{1.0, 0.21, 0.37}

\newcommand{\be}{\begin{equation}}
\newcommand{\ee}{\end{equation}}
\newcommand{\bea}{\begin{eqnarray}}
\newcommand{\eea}{\end{eqnarray}}
\usepackage{enumitem}

\newcommand\tcr{\textcolor{red}}

\newcommand\tcdg{\textcolor{darkgreen}}

\usepackage[most]{tcolorbox}

\tcbset{highlight math style={left=02mm,right=02mm,top=02mm,bottom=02mm}} 
\usepackage{empheq}

\begin{document}
\title{\tcr{An Inconsistency in the Null Strings Literature}: \tcdg{The Tale of an Overlooked Symmetry}}	

\author{M. M. Sheikh-Jabbari}
\email{jabbari@theory.ipm.ac.ir}
\affiliation{School of Physics, Institute for Research in Fundamental Sciences (IPM),
P.O. Box 19395-5531, Tehran, Iran}	
\author{Hossein Yavartanoo} 
\email{yavar@bimsa.cn}
\affiliation{Beijing Institute of Mathematical Sciences and Applications (BIMSA), Huairou District,
Beijing 101408, P.R. China}
\begin{abstract}
Null strings, strings with $2d$ null worldsheets, have about half a century of literature and are relevant in many physically interesting cases, especially when strings probe cosmological or black hole horizons. We observe that the null string action possesses a previously overlooked local symmetry, the consideration of which is necessary for the consistency of null string analyses. By correctly accounting for this symmetry, 
we show that the number of physical propagating degrees of freedom of null strings in $D$ dimensional target space is $D-3$, in contrast to  $D-2$ that one finds in the literature. In other words, null strings probe a codimension-1 null surface on a $D$ dimensional target space. The existence of this overlooked symmetry  calls for a thorough revision of  results in the null string literature. We briefly mention some of its physical consequences. 
\end{abstract}
		
\maketitle

String theory is proposed to be the high-energy completion of ordinary physics we are familiar with. It  comes with fundamental scale $l_s$, related to its tension $T$ as $l_s^2=1/(2\pi T)$. 
At energy scales $E \ll 1/l_s$, formally in $l_s \to 0$ limit, the theory is expected to reproduce standard quantum field theory  \cite{Polchinski:1998rq, Polchinski:1998rr}. The opposite limit, $E l_s\gg 1$, corresponds to ultra-high energy regimes near the Hagedorn transition \cite{Hagedorn:1965st} where the string features become very pronounced. This regime can be formally described by the ``tensionless'' limit $l_s \to \infty$ or $T \to 0$, also know as null string limit \cite{Schild:1976vq, Gamboa:1989zc, Isberg:1993av}. This limit is relevant for strings probing black hole or cosmological horizons where redshift factors diverge \cite{Lousto:1996hg, deVega:1994hu, Bagchi:2022iqb, Bagchi:2023cfp, Bagchi:2024rje}.

The worldsheet of tensile strings is a $1+1$ dimensional Lorentzian geometry described by the $2d$ metric defined on it and the Polyakov string theory is the most general theory one can write which has $2d$ diffeomorphisms and Weyl symmetry as local gauge symmetry \cite{Polchinski:1998rq, Polchinski:1998rr}. Worldsheet of tensile strings are hence a set of oscillators (vibrating string modes) subject to Virasoro constraints that are reminiscent of the worldsheet gauge symmetry. In contrast, in the tensionless null strings limit to have a well-defined worldsheet theory,  one is forced to replace the ordinary $2d$ worldsheet with a null surface, technically,  a $2d$ Carrollian geometry \cite{Bagchi:2021ban, Bagchi:2022nvj, Ciambelli:2025unn, Bagchi:2025vri}. 

A Carrollian worldsheet  is endowed with an extra Carroll-Weyl scaling symmetry, which has no counterpart in the tensile worldsheets \cite{Sheikh-Jabbari:2026vqh}. By the same token as in Polyakov string theory, the null string should be formulated as the theory that realizes this Carroll-Weyl scaling in addition to the tensile string gauge symmetries. From the target-space viewpoint, the null string worldsheet is a collection (congruence) of lightlike geodesics on the target-space; a collection of massless particles and hence probes a  a codimension-1 null surface. The peculiar feature of the null string that we uncover in this work arises directly from this fact. What renders the congruence into a string is the set of constraints resulting from the gauge symmetries of the theory. Taking the tensionless limit over the Virasoro constraints, one obtains BMS$_3$ constraints e.g. see \cite{Schild:1976vq,  Gamboa:1989zc,  Isberg:1993av, Lousto:1996hg, Bagchi:2013bga, Bagchi:2015nca, Bagchi:2022nvj} and also \cite{Lizzi:1989iy, Lizzi:1994rn, Cardona:2016ytk, Bagchi:2016yyf, Bagchi:2019cay, Bagchi:2020fpr, Bagchi:2020ats, Bagchi:2021rfw, Chen:2023esw, Bagchi:2024qsb}, see \cite{Bagchi:2026wcu} for a more complete set of references.

In this Letter, we demonstrate that  the Isberg, Lindstr\"om, Sundborg, and Theodoridis (ILST) action \cite{Isberg:1993av}, which is customarily viewed as the basic action defining null string theory, enjoys a larger gauge symmetry than those already discussed in the literature. Consequently, null strings are subject to an enlarged BMS$_3$ constraint algebra that we discuss. The role of this enlarged symmetry is to enforce the fact fact that null strings probe a codminesion-1 null surface in the target space. This overlooked symmetry and the resulting constraint calls for a thorough revision of the established 50 years of null strings analyses. This is meant to initiate this thorough revision.

\noindent{\textbf{$\blacksquare$ Null string action and the $\chi$-symmetry}}

$D$ dimensional null strings in  governed by  the ILST action \cite{Isberg:1993av}
\begin{equation}\label{NSA}
S = \frac{\kappa}{2} \int d^2\sigma \; \mathcal{V}^a \mathcal{V}^b \; \partial_a X^\mu \partial_b X_\mu,
\end{equation}
where $\mu=0,1,\cdots, D-1$, $a,b=1,2$, $\mathcal{V}^a$ is a worldsheet vector density and $\kappa$ is an arbitrary constant added to render the action dimensionless. The dynamical degrees of freedom in \eqref{NSA} are ${\mathcal{V}}^a$ and $X^\mu$ whose Euler-Lagrange equations of motion (EoM) are
\begin{subequations}\label{EoM-Ca-Xmu}
\begin{align}
    &\text{EoM for } {{\mathcal{V}}}^a:\qquad  C_a:={{\mathcal{V}}}^b \partial_a X^\mu \partial_b X_\mu=0 \label{Va-EoM}\\
   &\text{EoM for }  X^\mu:\qquad  \partial_a \left({{\mathcal{V}}}^a {{\mathcal{V}}}^b \partial_b X^\mu\right)=0  \label{Xmu-EoM}    
\end{align}
\end{subequations}
The action \eqref{NSA} is invariant under $2d$ diffeomorphisms $\sigma^a\to \sigma^a+\xi^a(\tau, \sigma)$, with
\begin{subequations}\begin{align}
    \delta_\xi X^\mu &= \xi^a \partial_a X^\mu, \label{diff_X} \\
    \delta_\xi \mathcal{V}^a &= \xi^b \partial_b \mathcal{V}^a - \mathcal{V}^b \partial_b \xi^a + \frac{1}{2} (\partial_b \xi^b) \mathcal{V}^a, \label{diff_V}
\end{align}
 \end{subequations} 
where the last term in \eqref{diff_V} accounts for the density weight $+1/2$ of $\mathcal{V}^a$.

Consider,  the $\chi$-transformation, 
\begin{equation}   \label{scaling_symmetry}
    \delta_\chi X^\mu = \chi X^\mu, \qquad \delta_\chi \mathcal{V}^a = -\chi \mathcal{V}^a,
\end{equation}
with $\chi=\chi(\tau,\sigma)$. Under the above \eqref{NSA}  transforms as
\begin{equation} \label{chi-variation}
    \delta_\chi S = \kappa \int d^2\sigma \; (\mathcal{V}^a \partial_a \chi)\ \mathcal{V}^b X^\mu \partial_b X_\mu.
\end{equation}
Off-shell invariance for arbitrary $X^\mu$ necessitates,
\begin{equation} \label{chi_constraint}
\mathcal{V}^a \partial_a \chi = 0.
\end{equation}
This restricts $\chi$ to a function of a single variable, rather than a generic function of $(\sigma, \tau)$. The ``$\chi$-symmetry,'' defined by \eqref{scaling_symmetry} subject to \eqref{chi_constraint}, is our main observation and has been overlooked in the null string literature. {Unlike a conventional Weyl symmetry with an arbitrary local parameter, the parameter $\chi$ is constrained by the Carrollian structure itself. However, one can show that this is reminiscent of a Carroll–Weyl scaling gauge symmetry, which, as discussed, should indeed be present in the null string theory \cite{Sheikh-Jabbari:2026vqh, Sheikh-Jabbari:2026tpf}}. The symmetry is therefore a local gauge redundancy with a restricted gauge parameter, analogous to residual gauge symmetries after gauge fixing.  Below, we discuss its profound consequences for  the consistency of the null string analyses. 

\paragraph{Temporal gauge fixing.} One can use diffeomorphisms to fix the ``temporal gauge'' and set the vector ${{\mathcal{V}}}^a$ to $(1,0)$ in $(\tau, \sigma)$ basis. The residual symmetries, i.e. a combination of diffeomorphisms+ scalings satisfying $\delta_\eta {{\mathcal{V}}}^a=0$, where $\eta=(\zeta^a, \chi)$, take the form,
\begin{equation}\label{resudual-sym}
    \zeta^a= \left[h(\sigma)+(f'(\sigma)-2\chi(\sigma))\tau\right] \partial_\tau + f(\sigma) \partial_\sigma, 
\end{equation}
where $f'=\partial_\sigma f$. {The residual gauge transformations are therefore parametrized by three functions of $\sigma$ only and}
\begin{equation}\label{X-under-residual}
   \delta_\eta X^\mu= \left[h+(f'-2\chi)\tau\right] \dot{X}^\mu+ f X^\mu{}'+\chi X^\mu,
\end{equation}
where $\dot{X}=\partial_\tau X$. The temporal gauge-fixing is guaranteed by imposing ${{\mathcal{V}}}^a$ EoM \eqref{Va-EoM} as constraints:
\begin{equation}\label{C1,C2-temporal}
    \text{C}_1:=P^\mu \dot{X}_\mu =0,\quad 
    \text{C}_2:=P^\mu X'_\mu =0, \quad P^\mu=\dot{X}^\mu
\end{equation}

\noindent{\textbf{$\blacksquare$ $\chi$-symmetry necessitates an extra constraint}}

To assess implications of the $\chi$-symmetry, consider the equations of motion in temporal gauge, $\ddot{X}^\mu=0$, and the constraints $\text{C}_1, \text{C}_2$ \eqref{C1,C2-temporal} yield
\begin{equation}
\begin{split}
    X^\mu(\tau,\sigma)=& P^\mu(\sigma) \tau+Q^\mu(\sigma),
\\
\mathcal{C}_1=P^2=0,&\qquad \mathcal{C}_2=P\cdot Q'=0 .
\end{split}
\end{equation}
Consistency of description then requires that under the residual symmetries generated by under the residual symmetry $\eta = (\zeta^a, \chi)$,  the constraints ${\cal C}_1, {\cal C}_2$ should be closed. One can then immediately verify that, 
\begin{equation}\label{Constraint-closure}
\begin{split}
\delta_\eta {\cal C}_1&=2(f'-\chi) {\cal C}_1 +f {\cal C}_1' \\ 
   \delta_\eta {\cal C}_2& = f {\cal C}'_2+ 2f' {\cal C}_2 +\tcr{\chi' {\cal C}_3}.
\end{split}
\end{equation}
where
\begin{equation}\label{C3-const}
{\cal C}_3:=P\cdot Q\ .
\end{equation}
Crucially, the variation $\delta_\eta {\cal C}_2$ contains the $\chi' {\cal C}_3$ term, highlighted in red; the closure of constraints requires existence of the third constraint ${\cal C}_3$. The above implies that the algebra is closed if ${\cal C}_3$ is a constant over the space of physical configurations. For completeness, we note that $ \delta_\eta {\cal C}_3= f {\cal C}'_3+ f' {\cal C}_3$ and hence the set of  ${\cal C}_1,  {\cal C}_2, {\cal C}_3$ constraints close under variations generated by $\eta$-transformations. 
Notice that closure of the algebra may be guaranteed, not through setting ${\cal C}_3$ to be a constant, but by requiring/restricting to $\chi' = 0$, e.g. as in  \cite{Isberg:1993av, Chen:2026klv}. Nonetheless, $\chi$ exists and there is no physical reason to restrict  to constant or vanishing $\chi$ sector; hence, necessitating the ${\cal C}_3$ constraint.

Given the importance of existence of the ${\cal C}_3$ constraint, we provide another argument for its existence, complementary to the closure of constraint algebra. To this end, we utilise the residual transformations \eqref{resudual-sym}, fix the standard light-cone gauge and solve for ${\cal C}_1, {\cal C}_2$. The details may be found in \cite{Sheikh-Jabbari:2026vqh, Sheikh-Jabbari:2026tpf}, here we sketch the argument. Passing to the light-cone gauge,
\begin{equation}\label{LCG-PX-plus}
P^+(\sigma)=p^+,\qquad Q^+(\sigma)=0,
\end{equation}
fixes two combinations of them:
\begin{equation}
h=0,\qquad \chi=f' ,
\end{equation}
and the ${\cal C}_1, {\cal C}_2$ constraints determine the longitudinal variables $P^-, Q^-$, leaving the transverse data
\begin{equation}
(Q^A(\sigma),P_A(\sigma)),\qquad A=1,\ldots,D-2 ,
\end{equation}

Due to the existence of the $\chi$-symmetry, we still have identification under $\chi$-transformations as well as the associated constraint. The remaining function $ f(\sigma)$  acts nontrivially as
\begin{equation}\label{residual-f}
\delta_f P^A=f P^{A\prime},\qquad \delta_f Q^A=(f x^A)' .
\end{equation}
This residual action is generated by
\begin{equation}
G[f]=-\oint d\sigma\,f\,Q^A P_A'=-\oint d\sigma\,f\, {\cal C}_3',
\end{equation}
where in the second equality we performed  an integration by-part, and used \eqref{LCG-PX-plus}. 
Thus the  $\chi$-symmetry, through the surviving parameter is $ \chi=f'$ in the light-cone gauge, requires
${\cal C}_3$= {constant for physical configurations}.  

\noindent{\textbf{$\blacksquare$ Implications for null string degrees of freedom}}

The physical phase space is therefore obtained by quotienting the $(Q^A, P_A)$ solution space by \eqref{residual-f} and requiring \eqref{C3-const}.  This removes one canonical pair from the usual $ D-2$  transverse
pairs $(Q^A, P_A)$. Thus, the number of local physical phase-space degrees of freedom is
\begin{equation}
{\rm dim.\ phys.\ phase\  space}=2(D-2)-2=2(D-3).
\end{equation}
Equivalently, a null string in $ D$  target dimensions carries $ D-3$  local
configuration-space degrees of freedom, not $ D-2$  as assumed in the standard
literature. The above analysis can be repeated more systematically using the standard method is the Dirac-Bergmann algorithm, see \cite{Henneaux:1992} and references therein. A more detailed analysis may be found in \cite{Sheikh-Jabbari:2026tpf}, reinforcing the degrees of freedom number count discussed above.

From a different perspective, as we discussed, null strings are a collection of null geodesics on the $D$ dimensional target space. As such, null strings can only probe a null $D-1$ dimensional hypersurface embedded in the $D$ dimensional target space. One can shows that \cite{NS-NT-progress} the role of $\chi$-symmetry and the resulting constraint ${\cal C}_3$ is to enforce this fact. Once restricted to this $D-1$ dimensional null surface, one needs only to enforce the standard ${\cal C}_1, {\cal C}_2$ (BMS$_3$) constraints. This is another argument for null strings possessing $D-3$ dynamical modes. 

\vskip 1mm
\centerline{\textbf{Concluding Remarks}}
\vskip 1mm

The $\chi$-symmetry defined by \eqref{scaling_symmetry}, \eqref{chi_constraint} has been systematically omitted from all prior analyses of the null string.  As we established the $\chi$-symmetry yields a third constraint ${\cal C}_3$ \eqref{C3-const} and an extra identification of $X^\mu$ and $e^{\chi} X^\mu$ \eqref{scaling_symmetry} in which $\chi$ is restricted to \eqref{chi_constraint}. This restricted scaling symmetry has no direct analogue in the usual tensile string formulation.  Stated more explicitly, \textit{the ILST action  describes a ``null string'' only when supplemented by the ${\cal C}_3$ constraint; without ${\cal C}_3$ it is a Carrollian linear $\sigma$-model, rather than a null string theory.}

As discussed in \cite{Sheikh-Jabbari:2026vqh}, the ILST action \eqref{NSA} may be viewed as the (partially) gauge-fixed form of an action that has $\chi(\tau, \sigma)$ as a gauge symmetry. This enhanced gauge symmetry is among the additional options available in Carrollian geometry (of the null string worldsheet) and has no counterpart in the textbooks of tensile strings. That is, this enhanced gauge symmetry is not visible once one obtains the null string action as a limit of the standard tensile string. 

The identification of $X^\mu$ and $e^{\chi} X^\mu$ breaks rigid target-space translations of $X^\mu \to X^\mu+ a^\mu$, which is an apparent global symmetry of the ILST action \eqref{NSA}. Nonetheless, the translations in $X^-$ direction and the constant value for ${\cal C}_3$ over the solution space are closely related to each other, once we recall that in the light-cone gauge adapted above, ${\cal C}_3= -p^+ Q^-+ P_A Q^A$. Since $p^+$ is a constant over the solution space, a constant shift in ${\cal C}_3$ can be associated with a rigid shift in $X^-$ (or equivalently a rigid shift in $Q^-$). In other words, one can always set ${\cal C}_3=0$ on the solution space by the choice of $X^-$ coordinate origin. 

Eq.~\eqref{Constraint-closure} and $\delta_\eta {\cal C}_a$ can be expressed more conveniently in terms of Poisson brackets among the constraints ${\cal C}_1, {\cal C}_2$, and ${\cal C}_3$. Denoting the Fourier modes of ${\cal C}_1$, ${\cal C}_2$, and ${\cal C}_3$ by $L_n$, $M_n$, and $S_n$, respectively, we obtain the algebra
\begin{subequations}
\begin{align}
[L_n, L_m] &= (n-m)L_{n+m}, \\ 
[L_n, M_m] &= (n-m)M_{n+m}, \label{eq:BMS} \\
[L_n, S_m] &= -m S_{n+m}, \\ 
[S_n, M_m] &= 2M_{n+m}, \label{eq:WBMS}
\end{align}
\end{subequations}
with all other commutators vanishing. This algebra has previously appeared in the literature~\cite{Adami:2020ugu,Adami:2021nnf,Batlle:2024ahz} and is the BMS$_3$ algebra (generated by $L_n$ and $M_n$) enlarged by the $S_n$ generators, which transform as a spin-$1$ field under BMS$_3$. 

The $\chi$-symmetry is a mandatory component of the null string's gauge structure and its proper treatment  is necessary for the consistency of description. Its has crucial, yet to be studied, implications for null string and  its quantization, mode expansions, interactions. Preliminary results on quantized null string theory  indicate that it may have a discrete spectrum \cite{Rasulian:2026jvg}. This and its many other  physical implications of the $\chi$-symmetry, e.g. for strings probing cosmological and/or black hole horizons and for  null boundary holography constitute the next steps of the null string theory research program. 


\begin{acknowledgments}
We thank Arjun Bagchi, Aritra Banerjee, Ulf Danielson, Daniel Grumiller, Ulf Lindstr\"om, Ida Rasulian, Bo Sundberg  and Shing-Tung Yau for discussions. MMShJ acknowledges Iranian National Science Foundation (INSF) research chair grant No.40451653. HY is supported in part by Beijing Natural Science Foundation under Grant No. IS23013.
\end{acknowledgments}


\end{document}